\documentclass[nofootinbib,prd,reprint,superscriptaddress,preprintnumbers,longbibliography,floatfix]{revtex4-2}
\usepackage{float}
\usepackage[bottom]{footmisc}
\usepackage{amsmath,amssymb}
\usepackage{graphicx}
\usepackage{bm}
\usepackage{xcolor}
\usepackage[colorlinks=true, allcolors=blue]{hyperref}

\begin{document}

\preprint{LA-UR-26-28779}

\title{Neutrino flavor instabilities intermediate between slow and fast}

\author{Nicol\'{a}s Viaux}
\affiliation{Departamento de F\'{i}sica, Universidad T\'{e}cnica Federico Santa Mar\'{i}a, Casilla 110-V, Valpara\'{i}so, Chile}
\email{nicolas.viaux@usm.cl}

\author{Lucas Johns}
\affiliation{Theoretical Division, Los Alamos National Laboratory, Los Alamos, NM 87545, USA}
\email{ljohns@lanl.gov}

\begin{abstract}
The neutrino flavor composition in supernovae and neutron star mergers is shaped by flavor instabilities, which are commonly classified into three types: slow (SFI), fast (FFI), and collisional (CFI). The focus here is on mixed-type instabilities combining attributes of the standard varieties. We examine instability growth rates in two-dimensional parameter spaces (``phase diagrams'') whose axes correspond to instabilities of pure type. We find critical curves connecting the pure-type instability onset points. We also exhibit qualitative trends in the growth-rate spectra $\gamma_{\bm{k}}$, emphasizing the occurrence of crossovers and phase transitions in the fastest-growing Fourier mode $k_{\textrm{max}}$ and drawing attention to the distinction between spatial-mean ($\bm{k} = 0$) and flavor-wave ($\bm{k} \neq 0$) instabilities. Our paper is a step toward understanding flavor instabilities as they actually occur in nature rather than in idealized limits.
\end{abstract}

\maketitle

\section{Introduction}

Dense neutrino gases in core-collapse supernovae and neutron-star mergers can convert flavor collectively through neutrino-neutrino forward scattering~\cite{pantaleone1998stability,duan2006coherent,duan2006simulation,duan2010collective,mirizzi2015,chakraborty2016collective,tamborra2021new,volpe2024neutrinos,johns2025neutrino}. The mean-field quantum kinetic equation (QKE) is the standard setting, and linear stability analysis of that equation is the standard diagnostic of instabilities~\cite{banerjee2011linearized,sarikas2012,yi2019dispersion,fiorillo2024theory}. Within that framework the literature has organized conversion into three pure types---slow, fast, and collisional~\cite{johns2021collisional}---and has often separated those instabilities and worked in idealized limits for linear stability analysis~\cite{akaho2024collisional,liu2023universality}. This paper maps the transition between the first two.

Slow flavor instabilities (SFI) were the first collective mode to be recognized. They are driven by the vacuum oscillation frequency $\omega=\Delta m^2/(2E)$, with $\Delta m^2$ the mass-squared splitting and $E$ the neutrino energy, together with a crossing of the energy-dependent lepton-number spectrum~\cite{kostelecky1993neutrino,hannestad2006self,duan2007analysis,raffelt2007self,dasgupta2008split,dasgupta2009multiple,dasgupta2010hierarchy,johns2018strange,fiorillo2023slow,fiorillo2025theoryslow,fiorillo2025theoryslow2}. The classic analyses are homogeneous or nearly so: a bipolar pendulum, a spectral swap, an instability that disappears if $\omega$ is set to zero~\cite{mangano2014damping,dasgupta2015temporal}. For a long time this was the working picture of collective oscillations in supernovae.

Fast flavor instabilities (FFI) shifted the emphasis. Sawyer~\cite{sawyer2005speed,sawyer2016neutrino} showed that an angular instability can grow on a scale set by the neutrino-neutrino coupling alone, even in the formal limit $\omega\to 0$. That observation produced a large literature in which the vacuum frequency is dropped, the electron-lepton-number (ELN) angular distribution is the control parameter, and an ELN crossing is the diagnostic~\cite{izaguirre2017fast,chakraborty2016self,chakraborty2016collective,capozzi2017fast,capozzi2019fast,morinaga2020fast,morinaga2022fast,dasgupta2022collective,johns2025implications,abbar2019occurrence,abbar2021characteristics,nagakura2021where,delfan2019linear,johns2020neutrino,johns2020fast,johns2021crossings,nagakura2021moments}. In that limit a crossing is necessary for FFI~\cite{morinaga2022fast,dasgupta2022collective,johns2025implications}. The same focus on FFI underpins much of the phenomenological discussion of flavor conversion in supernovae and in neutron-star mergers~\cite{wu2017fast,li2021neutrino,george2020fast,grohs2023neutrino}: one asks whether the local angular distributions are crossed, adopting the FFI limit.

The result is a field that knows the pure types well and treats SFI and FFI as alternatives. A paper on slow conversion sets $\omega$ finite and the angular distributions nearly isotropic. A paper on fast conversion sets $\omega=0$ and scans the ELN. Each is an idealization: in the dense interior of a supernova or merger the vacuum frequency and the angular anisotropy are finite at once~\cite{tamborra2021new,shalgar2023neutrino,johns2025neutrino}. Mixed-type instabilities exist between those two limits.

Recent work has combined fast and slow ingredients~\cite{dasgupta2025sufficient,fiorillo2025dispersion}; those papers address the existence of unstable modes when energy and angle both enter a crossing of the lepton-number distribution. This paper maps mixed-type instabilities away from the idealized slow and fast limits. An instability that still grows as $\omega\to 0$ is fast in the sense of Sawyer~\cite{sawyer2005speed,sawyer2016neutrino}. Slow instabilities can be seen as modes that disappear at $\omega=0$, or as modes that persist when the angular distributions are restored to isotropy. Away from the $\omega=0$ and isotropic limits, SFI and FFI are blended together.

The aim, concretely, is a linear analysis of the SFI--FFI transition in a two-dimensional parameter space whose axes correspond to the two pure types, followed by a collisionless nonlinear check of both sides of the wall. Section~\ref{sec:methods} writes the two-flavor mean-field QKE and the linearized problem, with every symbol defined in the order it enters. One axis turns on angular FFI, the other turns on vacuum-frequency SFI. Instabilities away from those axes mix qualities of both types; they can be projected onto the idealized limits by setting $\omega=0$, by isotropizing the angular distributions, or both. A Fermi--Dirac family in Sec.~\ref{sec:fd} checks that the transition is not an artifact of the ramp occupations. Section~\ref{sec:nl} evolves the two sides of the wall in a one-dimensional periodic box with $N_x=512$: SFI saturates as a near-homogeneous mode, FFI as a flavor wave at the linear $k_*$~\cite{johns2020fast,richers2022code,wu2021collective,duan2021flavor}. Collisional flavor instabilities (CFI), the third pure type~\cite{johns2021collisional}, are left aside. The paper is a step toward flavor instabilities as they occur in compact-object environments rather than in the isolated limits that have organized the literature.

\section{Methodology}
\label{sec:methods}

The two-flavor mean-field QKE is the equation of motion used throughout. Each discrete mode $i$ is a pair $(v_i,\omega_i)$, with velocity cosine $v_i=\cos\theta_i\in[-1,1]$ and vacuum frequency $\omega_i=\Delta m^2/(2E_i)>0$. Neutrinos and antineutrinos are both present; we label the species of mode $i$ by $\varepsilon_i=+1$ ($\nu$) or $\varepsilon_i=-1$ ($\bar\nu$). The flavor content of the mode is a $2\times 2$ Hermitian density matrix $\rho_i$ in the $(e,x)$ basis.

The Hamiltonian of that mode is vacuum plus neutrino-neutrino mean field. Let $\theta_v$ be the vacuum mixing angle, $\sigma_x$ and $\sigma_z$ the Pauli matrices, $G_F$ the Fermi constant, $n_\nu$ the local neutrino number density, and $\mu$ the neutrino-neutrino coupling (proportional to $\sqrt{2}\,G_F n_\nu$). Then
\begin{equation}
  H_i
  = H_{\mathrm{vac}}(\varepsilon_i\omega_i,\theta_v)
  + \mu\sum_j (1-v_i v_j)\,\rho_j,
  \label{eq:ham}
\end{equation}
\begin{equation}
  H_{\mathrm{vac}}(\omega,\theta_v)
  =\frac{\omega}{2}\bigl(-\cos 2\theta_v\,\sigma_z+\sin 2\theta_v\,\sigma_x\bigr).
  \label{eq:hvac}
\end{equation}
The factor $\varepsilon_i$ in Eq.~(\ref{eq:ham}) implements the opposite vacuum frequency of antineutrinos. Because $\omega_i>0$ and the $\sigma_z$ coefficient in Eq.~(\ref{eq:hvac}) is $-\cos 2\theta_v$, this convention corresponds to normal mass ordering ($\Delta m^2>0$); the inverted ordering flips that sign, which alters the SFI portion of the phase diagrams while leaving the $\omega\to 0$ (FFI) limit unchanged. The mean-field sum in Eq.~(\ref{eq:ham}) runs over both species with no explicit sign because we use the flavor-isospin convention: for antineutrino modes $\rho_j$ carries a negative occupancy [Eq.~(\ref{eq:background}) below], which automatically weights antineutrinos with the opposite sign in the mean field. The current-current kernel $1-v_i v_j$ is the sum of a monopole and a dipole; that rank-2 structure is what closes the linearized problem on a $2\times 2$ matrix rather than on a dense $N\times N$ matrix~\cite{banerjee2011linearized,sarikas2012,yi2019dispersion}. Matter is off: no charged-current potential is added to $H_i$. The collisional term is off. Along a one-dimensional ray $z$, $\rho_i$ obeys
\begin{equation}
  (\partial_t + v_i\partial_z)\,\rho_i
  = -i\bigl[H_i,\rho_i\bigr].
  \label{eq:qke}
\end{equation}

Each mode carries a signed occupancy $g_i$, the flavor-isospin lepton number of that bin including its angular quadrature weight $w_i$. The velocity cosines $v_i$ are uniformly spaced on $[-1,1]$. The weights implement the trapezoidal rule for $\int_{-1}^{1} dv/2$,
\begin{equation}
  w_i
  =
  \begin{cases}
    1/[2(N_v-1)] & i=1,N_v,\\
    1/(N_v-1) & \text{otherwise},
  \end{cases}
  \label{eq:wi}
\end{equation}
so that $\sum_i w_i=1$. The quantity $g_i$ is the flavor-isospin difference $n_e-n_x$ of that bin (with a minus sign for antineutrinos), not the physical $e$ occupation. On a flavor-diagonal background the coherences vanish, $\rho_{i,ex}^{(0)}=0$, and the linearized kernel depends only on that difference. We store it as
\begin{equation}
  \rho_{i,ee}^{(0)}=g_i,
  \qquad
  \rho_{i,xx}^{(0)}=0,
  \qquad
  \rho_{i,ex}^{(0)}=0,
  \label{eq:background}
\end{equation}
which is equivalent, for linear stability, to physical occupations $\rho_{i,ee}^{(0)}=n_e$ and $\rho_{i,xx}^{(0)}=n_x$. The ramp family below has $n_x=0$; the Fermi--Dirac family of Sec.~\ref{sec:fd} does not.
The monopole and dipole of that background are
\begin{equation}
  G_0\equiv\sum_j g_j,
  \qquad
  G_1\equiv\sum_j v_j g_j.
  \label{eq:moments}
\end{equation}

Linear stability is the spectrum of small flavor coherences on top of Eq.~(\ref{eq:background}). Let $k$ be the wavenumber along $z$ and $\lambda\in\mathbb{C}$ the complex eigenfrequency of the linearized QKE. A Fourier mode is then
\begin{equation}
  \delta\rho_i(z,t)
  =\delta\tilde\rho_i\,e^{\lambda t+ikz}+\mathrm{c.c.}
  \label{eq:ansatz}
\end{equation}
where $\delta\tilde\rho_i$ is the $2\times 2$ Fourier amplitude and $\mathrm{c.c.}$ denotes the complex conjugate. The quantity $\lambda$ is defined completely by this ansatz: it is an eigenvalue of that linearized problem. We decompose it as
\begin{equation}
  \lambda=\gamma+i\Omega,
  \label{eq:lambda}
\end{equation}
with growth rate $\gamma\equiv\mathrm{Re}\,\lambda$ and oscillation frequency $\Omega\equiv\mathrm{Im}\,\lambda$. A mode with $\gamma>0$ grows exponentially; a mode with $\gamma=0$ is neutrally oscillating; a mode with $\gamma<0$ is decaying and therefore ignored in our analysis: we keep $\gamma\ge 0$ only. If $J$ denotes the Jacobian of the linearized QKE, so that $\partial_t\delta\rho=J\,\delta\rho$, then $\lambda$ is an eigenvalue of $J$. That is not the $e^{-i\Omega t}$ convention in which growth is sometimes written as an imaginary part of $\Omega$.

Linear stability analysis is performed on a flavor-diagonal background, $\rho_{i,ex}^{(0)}=0$. The vacuum mixing angle is $\theta_v=10^{-6}$. Only the diagonal vacuum precession $\cos 2\theta_v$ enters the linearized kernel through $D_i$ in Eq.~(\ref{eq:Di}) below; at this $\theta_v$ one has $\cos 2\theta_v\simeq 1$. The off-diagonal vacuum piece $(\omega/2)\sin 2\theta_v\,\sigma_x$ in Eq.~(\ref{eq:hvac}) is a seed for the nonlinear evolution of Sec.~\ref{sec:nl}. It does not enter the linearized problem and does not shift $\gamma$ on the maps.

Linearizing Eq.~(\ref{eq:qke}) in the Fourier amplitudes of the coherences $S_i\equiv\delta\tilde\rho_{i,ex}$ produces a closed system. The kinematic-plus-mean-field precession of mode $i$ at wavenumber $k$ is
\begin{equation}
  D_i(k)
  =-\varepsilon_i\omega_i\cos 2\theta_v
  +\mu\bigl(G_0-v_i G_1\bigr)
  +k v_i.
  \label{eq:Di}
\end{equation}
The three terms are vacuum precession (opposite for $\bar\nu$), the flavor-isospin mean field of the background, and spatial advection of the Fourier mode. Every later growth rate is a root whose poles sit at $\lambda=-i D_i(k)$. Because $1-v_i v_j$ is rank-2, the angular moments of $S_i$ collapse to the three sums
\begin{equation}
  I_n(\lambda;k)
  =\sum_i\frac{g_i\,v_i^n}{\lambda+i D_i(k)},
  \qquad n=0,1,2.
  \label{eq:In}
\end{equation}
$I_0$, $I_1$, and $I_2$ are the occupancy-weighted monopole, dipole, and quadrupole of the coherences, each carrying the propagator $1/(\lambda+i D_i)$. A value of $\lambda$ is an eigenfrequency of wavenumber $k$ if and only if it is a root of the $2\times 2$ determinant built from those three moments,
\begin{equation}
  \det
  \begin{pmatrix}
    1-i\mu I_0(\lambda;k) & i\mu I_1(\lambda;k) \\
    -i\mu I_1(\lambda;k) & 1+i\mu I_2(\lambda;k)
  \end{pmatrix}
  =0.
  \label{eq:disp}
\end{equation}
All maps and spectra below are determined from this equation. At fixed $k$ we take the root of Eq.~(\ref{eq:disp}) of largest real part; the growth rate of that wavenumber is
\begin{equation}
  \gamma_k\equiv\mathrm{Re}\,\lambda(k),
  \label{eq:gammak}
\end{equation}
with $\lambda(k)$ that leading root. The two order parameters of the maps are the maximum growth rate and the wavenumber at which it occurs,
\begin{equation}
  \gamma_{\max}
  \equiv\max_k\gamma_k,
  \qquad
  k_*
  \equiv\arg\max_k\gamma_k,
  \label{eq:gmaxkstar}
\end{equation}
evaluated on a uniform grid $k/\mu\in[-8,8]$. The reported $(k_*,\gamma_{\max})$ is then refined by fitting a parabola through the three grid points around the discrete maximum of $\gamma_k$, so $k_*$ is not locked to a grid node. The homogeneous rate is the $k=0$ slice,
\begin{equation}
  \gamma_0\equiv\gamma_{k=0}.
  \label{eq:gamma0}
\end{equation}
A growing $\gamma_0$ means the spatially uniform background itself is unstable, as opposed to a flavor wave at $k\neq 0$.

The first family of backgrounds is an angular ramp. Occupation numbers (before quadrature weights) are independent of energy,
\begin{align}
  n_\nu(v)
  &= n_\nu^{(0)}
  +\frac12\eta_F(v-v_\times),
  \label{eq:nnu}\\
  n_{\bar\nu}(v)
  &= n_{\bar\nu}^{(0)}
  -\frac12\eta_F(v-v_\times),
  \label{eq:nnubar}
\end{align}
with crossing cosine $v_\times=1/2$. Baselines are $n_\nu^{(0)}=1$ and $n_{\bar\nu}^{(0)}=0.2$; the heavy-lepton occupation is $n_x=0$ on this family. The angular control parameter $\eta_F$ makes the distributions anisotropic in opposite directions for $\nu$ and $\bar\nu$. Occupancy positivity on $v\in[-1,1]$ requires $\eta_F\le 0.8$. Signed occupancies in Eq.~(\ref{eq:background}) are then
\begin{equation}
  g_i
  =
  \begin{cases}
    +\,n_\nu(v_i)\,w_i & \text{if $\varepsilon_i=+1$},\\
    -\,n_{\bar\nu}(v_i)\,w_i & \text{if $\varepsilon_i=-1$},
  \end{cases}
  \label{eq:gi}
\end{equation}
with $w_i$ the trapezoidal weights of Eq.~(\ref{eq:wi}). The energy grid is uniform, $\omega_j\in[\omega_{\min},\omega_{\max}]$ with $\omega_{\min}=10^{-3}\,\omega_{\max}$, and is used only to set the vacuum frequencies. When the energy grid $N_\omega$ is changed, every $g_i$ is multiplied by $8/N_\omega$ so that the total lepton number is held fixed against the change of binning.

\begin{figure}[htbp]
    \centering
    \includegraphics[width=\linewidth]{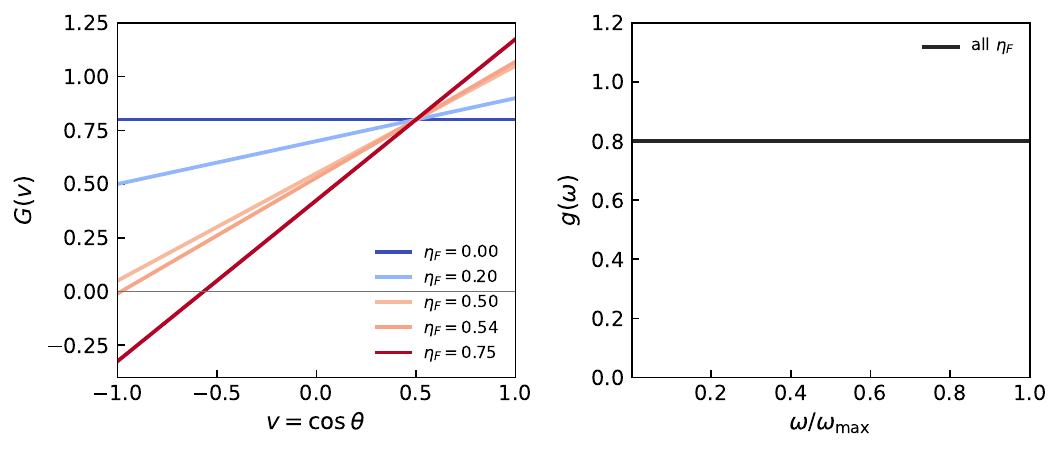}
    \caption{Ramp lepton-number distributions at $\eta_F=0.00$, $0.20$, $0.50$, $0.54$, and $0.75$. Left: energy-integrated ELN $G(v)$ of Eq.~(\ref{eq:Gv}). A crossing appears for $\eta_F>8/15$. Right: angle-integrated spectral lepton number $g(\omega)$. Occupations are independent of energy, so $g(\omega)$ is flat and the same for every $\eta_F$.}
    \label{fig:dist_ramp}
\end{figure}

The remaining control parameter is the vacuum-to-self-interaction ratio
\begin{equation}
  s\equiv\frac{\omega_{\max}}{\mu}.
  \label{eq:s}
\end{equation}
All growth rates and wavenumbers below are in units of $\mu$. Implementing $s$ by scaling the vacuum spectrum at fixed $\mu$, or by scaling $\mu$ at fixed $\omega_{\max}$, is therefore equivalent for the maps. Small $s$ is the dense interior; $s\sim 1$ is the dilute regime, where vacuum frequencies are not negligible. The energy-integrated ELN
\begin{equation}
  G(v)
  =\int d\omega\,\bigl[n_\nu(v)-n_{\bar\nu}(v)\bigr]
  \label{eq:Gv}
\end{equation}
is proportional to $0.8+\eta_F(v-v_\times)$ and therefore has a zero within $v\in[-1,1]$ if and only if $\eta_F>8/15$. The crossing sits at $v_c=v_\times-0.8/\eta_F$, entering the domain at $v=-1$ when $\eta_F=8/15$ and moving toward $v_\times$ as $\eta_F$ grows. Thus, at $\omega=0$, we expect FFI at $\eta_F>8/15$~\cite{morinaga2022fast,dasgupta2022collective,johns2025implications}. Figure~\ref{fig:dist_ramp} shows the energy-integrated ELN $G(v)$ and the angle-integrated spectral lepton number $g(\omega)$ for $\eta_F=0.00$, $0.20$, $0.50$, $0.54$, and $0.75$. Because the occupations of Eqs.~(\ref{eq:nnu})--(\ref{eq:nnubar}) are independent of energy, $g(\omega)$ is flat and independent of $\eta_F$.

The maps of the next section are drawn in the order parameters $\gamma_{\max}$ and $k_*$ of Eq.~(\ref{eq:gmaxkstar}). Which path a compact-object neutrino gas traces through those maps depends on the astrophysical driving: density sets $s=\omega_{\max}/\mu$, while the angular distributions set which species is more forward-peaked. On this family the angular FFI dial is $\eta_F$ and the vacuum-frequency dial is $s$.

\section{The slow--fast transition}
\label{sec:transition}

The physical space of neutrino distributions is infinite-dimensional. A reduced two-dimensional parameter space is most useful if each axis is a pure-type dial. Eqs.~(\ref{eq:nnu})--(\ref{eq:s}) give that space: $\eta_F$ turns on an ELN angular crossing and nothing else, while $s$ turns on vacuum frequency relative to $\mu$ and nothing else. Pure FFI occurs on the line $s\to 0$ with $\eta_F>8/15$. Pure SFI occurs on the line $\eta_F=0$ with $s$ of order unity. In the interior both knobs are on; the type is still unique, read from whether $\gamma_{\max}$ survives $s\to 0$.

\begin{figure}[htbp]
    \centering
    \includegraphics[width=.75\linewidth]{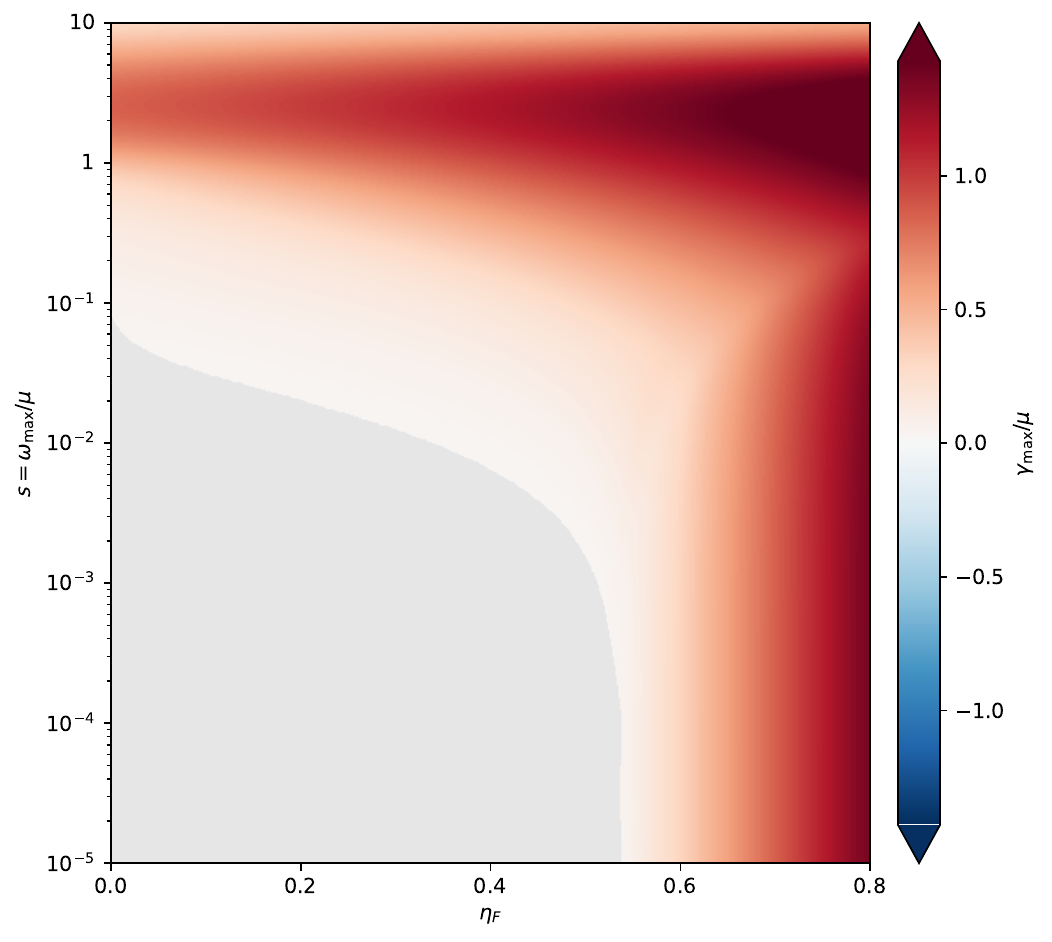}
    \caption{Maximum growth rate $\gamma_{\max}/\mu$ of Eq.~(\ref{eq:gmaxkstar}) on the ramp family, $N_v=N_\omega=256$. Axes $(\eta_F,s)$ are the angular and vacuum-frequency dials of Eqs.~(\ref{eq:nnu}) and~(\ref{eq:s}). Grey: numerical floor ($\gamma_{\max}/\mu<0.02$). For $\eta_F\lesssim 8/15$ a growing mode at large $s$ is gone as $s\to 0$. For $\eta_F\gtrsim 8/15$ growth continues to the bottom of the plot.}
    \label{fig:gmax_FFISFI}
\end{figure}

Figure~\ref{fig:gmax_FFISFI} is the maximum growth rate $\gamma_{\max}(\eta_F,s)$ of Eq.~(\ref{eq:gmaxkstar}). The horizontal axis is the angular FFI dial $\eta_F$. The vertical axis is $s=\omega_{\max}/\mu$ on a logarithmic scale. Resolution is $N_v=N_\omega=256$. Grey is a numerical floor, not a third phase: no growing root of Eq.~(\ref{eq:disp}) is trusted if $\gamma_{\max}/\mu<0.02$.

The classification is the vertical structure of this figure, not a color. For $\eta_F\lesssim 8/15$, left of the ELN crossing of $G(v)$, there is a growing mode at $s\sim 1$ and none (grey) as $s\to 0$. Those backgrounds need vacuum frequency, as in the classic slow analyses~\cite{hannestad2006self,fiorillo2023slow}. For $\eta_F\gtrsim 8/15$ a wall of growth continues to the bottom of the plot. Those backgrounds still grow when $\omega_{\max}$ is removed: they are FFI in the sense of Sawyer~\cite{sawyer2005speed,sawyer2016neutrino,izaguirre2017fast}. The onset of the wall tracks the crossing of $G(v)$, as the $\omega=0$ theorems require~\cite{morinaga2022fast,dasgupta2022collective,johns2025implications}. At finite $s$ both sides of the crossing can be unstable; the wall is the part that remains as $s\to 0$. The color is $\gamma_{\max}/\mu$, not $\gamma_{\max}$: a large value at large $s$ is not a larger dimensional growth rate. It is a linear rate, not a conversion depth.

\begin{figure}[htbp]
    \centering
    \includegraphics[width=.75\linewidth]{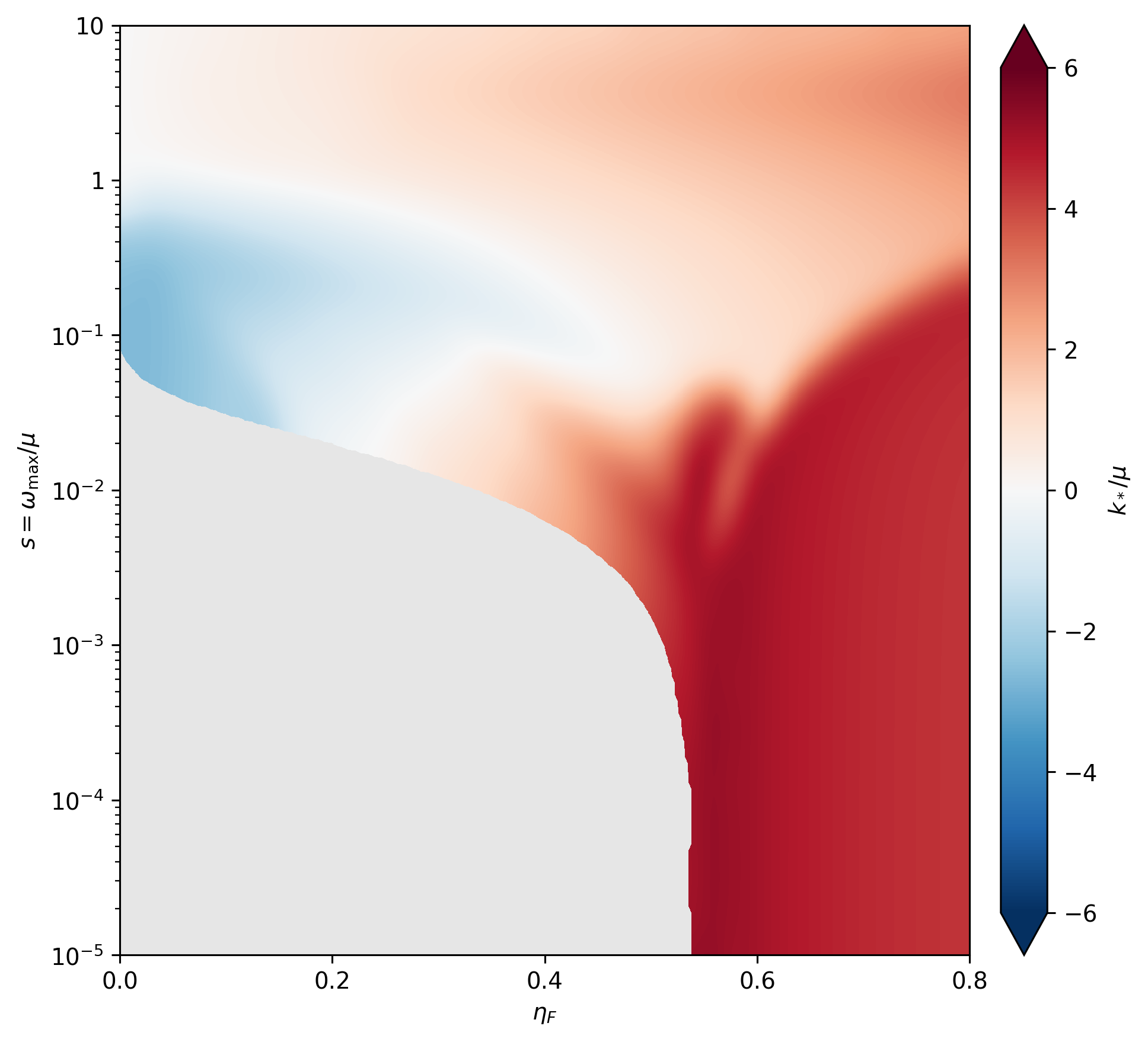}
    \caption{Wavenumber $k_*/\mu$ of the fastest-growing mode, Eq.~(\ref{eq:gmaxkstar}), on the same ramp plane as Fig.~\ref{fig:gmax_FFISFI}. Color is the sign of $k_*$, which follows which species is more forward-peaked, not SFI versus FFI. Red: $k_*>0$; blue: $k_*<0$; white: $k_*\approx 0$. Grey: no growing mode.}
    \label{fig:kstar_FFISFI}
\end{figure}

The same plane colored by $k_*/\mu$ is Fig.~\ref{fig:kstar_FFISFI}. The sign of $k_*$ follows which species is more forward-peaked. It is the sign of the wavenumber of fastest growth, not the direction of travel (that is the group velocity) and not an SFI/FFI classifier. The red wall at large $\eta_F$ is $k_*>0$ (neutrinos more forward than antineutrinos). The blue lobe at small $\eta_F$ and $s\sim 0.05$--$1$ is $k_*<0$. White is $k_*\approx 0$, a spatially uniform mode. Comparing the two figures: the FFI wall of Fig.~\ref{fig:gmax_FFISFI} is the high-$|k_*|$ red region that survives $s\to 0$; the SFI-only region sits at small $|k_*|$. A crossover of $k_*$ at fixed $s$ when $\eta_F$ crosses $\sim 8/15$ is the spectral signature of the same wall: the location of the maximum of $\gamma_k$ moves from near $k=0$ to finite $k$. That is the content of the next figure.

\begin{figure*}
    \centering
    \includegraphics[width=\linewidth,height=0.42\textheight,keepaspectratio]{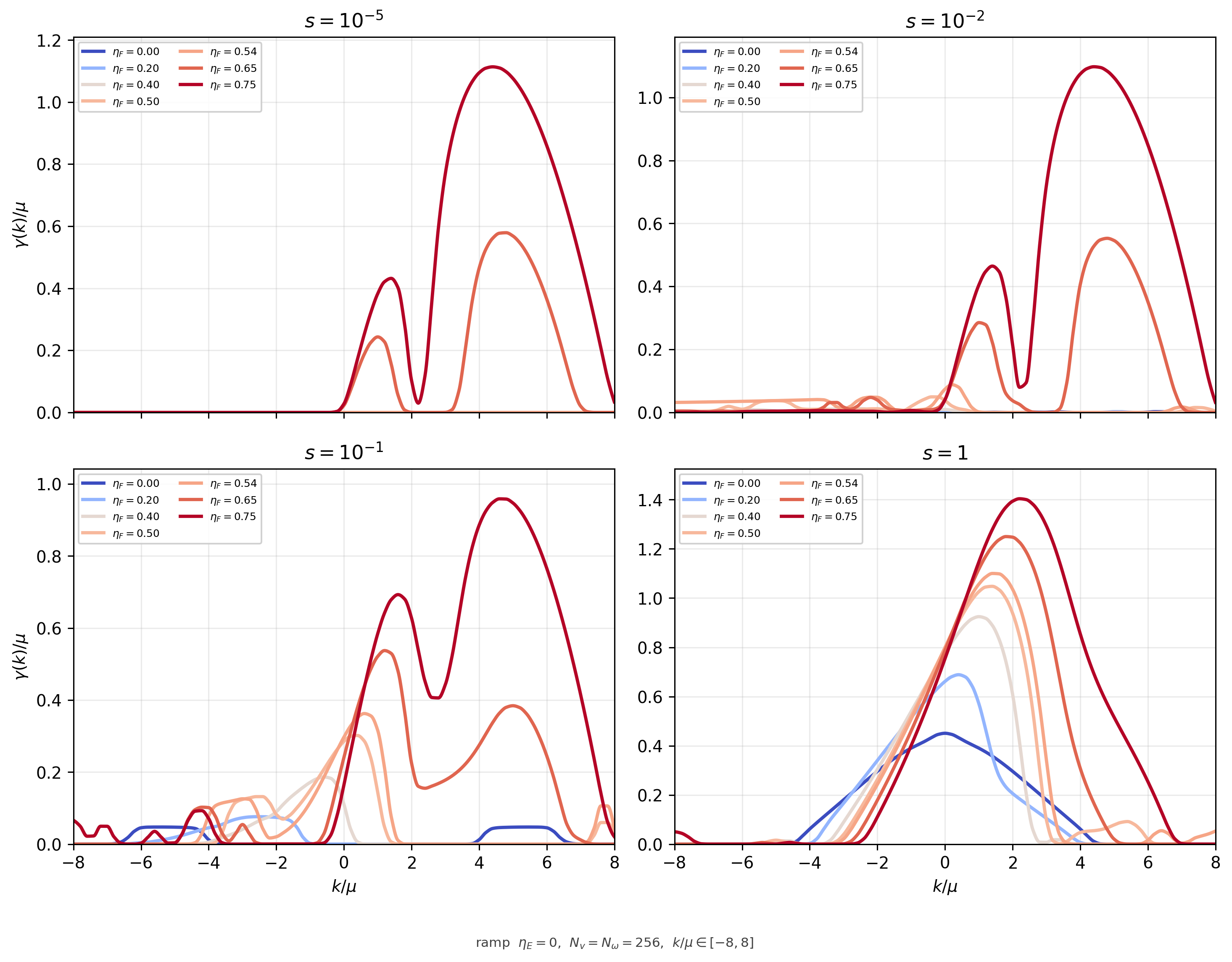}
    \caption{Growth-rate spectra $\gamma_k$ of Eq.~(\ref{eq:gammak}) versus $k$ for the ramp, $N_v=N_\omega=256$. Each panel is one $s$; curves are $\eta_F$. At small $s$ the growing spectra are peaked at finite $k$, with $\gamma_0\approx 0$. At $s=1$ they have unimodal character even at $\eta_F>0$.}
    \label{fig:gammak_FFISFI}
\end{figure*}

Figure~\ref{fig:gammak_FFISFI} is the spectrum $\gamma_k$ of Eq.~(\ref{eq:gammak}) at four values of $s$, with curves labeled by $\eta_F$. Both heatmaps are summaries of this object: $\gamma_{\max}$ is its maximum and $k_*$ is the $k$ at which that maximum is attained.

At $s=10^{-5}$ only $\eta_F\gtrsim 8/15$ grows. The spectrum is bimodal and peaked at finite $k\sim 4$--$6$, with $\gamma_0\approx 0$, as in the $\omega\to 0$ analyses of FFI~\cite{sawyer2016neutrino,chakraborty2016self,yi2019dispersion,duan2021flavor}. Small $\eta_F$ is identically $\gamma_k=0$, which is the grey of Fig.~\ref{fig:gmax_FFISFI}.

At $s=10^{-2}$ the same FFI peaks remain. A weaker growing branch appears at $k<0$ for intermediate $\eta_F$; that branch is the blue lobe of Fig.~\ref{fig:kstar_FFISFI}.

At $s=10^{-1}$ the curves with $\eta_F=0.00$--$0.50$ turn on as peaks near $k=0$. On the wall the finite-$k$ peak still wins, so $k_*$ remains large.

At $s=1$ every $\eta_F$ in the scan is unstable. The spectra have unimodal character even at $\eta_F>0$. The wavenumber $k_*$ can still undergo a smooth crossover as $\eta_F$ or $s$ is varied, while $\gamma_{\max}$ itself remains smooth.

\section{More realistic distributions}
\label{sec:fd}

The linearized problem is unchanged: Eqs.~(\ref{eq:qke})--(\ref{eq:gmaxkstar}) still define $\lambda$, $\gamma_k$, $\gamma_{\max}$, and $k_*$. What changes is the background $g_i$. Occupations are Minerbo functions~\cite{minerbo1978}
\begin{equation}
  f(E,v)
  =\bigl(1+e^{\beta\hat{E}-\alpha v}\bigr)^{-1},
  \label{eq:minerbo}
\end{equation}
with dimensionless energy $\hat{E}=E/E_0$ in units of a fixed reference energy $E_0$, and a spectral-hardness parameter $\beta$ that is independent of that choice of units. Different species have different $\beta$ on the same $\hat{E}$ grid, so $\beta$ is not absorbed into $\hat{E}$. The flux parameter $\alpha$ is the angular dial: $\alpha>0$ is forward-peaked, $\alpha=0$ is isotropic. Vacuum frequencies on this family are $\omega_j/\mu=s/\hat{E}_j$, with $s$ playing the same role as in Eq.~(\ref{eq:s}); on this family $s=\omega(\hat{E}=1)/\mu$ sets the overall vacuum-to-self-interaction ratio, coinciding with $\omega_{\max}/\mu$ only if the energy grid begins at $\hat{E}=1$.

\begin{figure}[htbp]
    \centering
    \includegraphics[width=\linewidth]{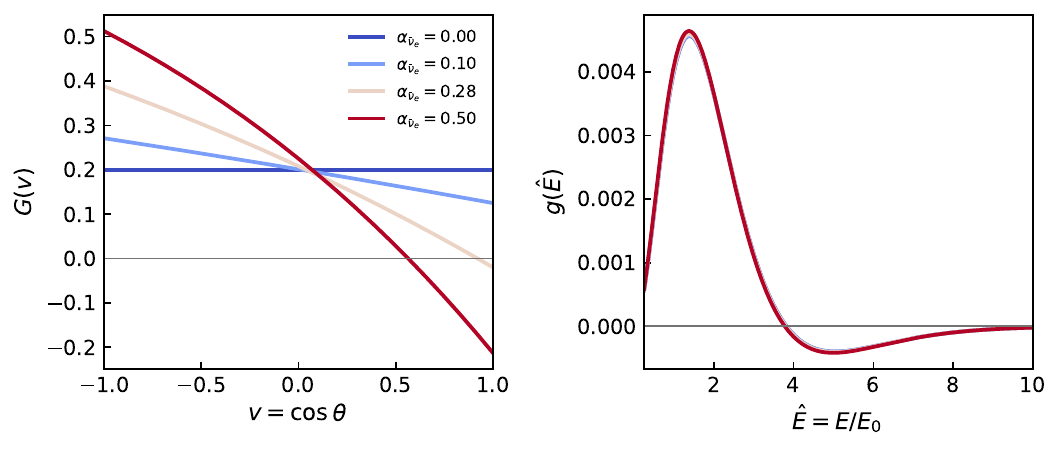}
    \caption{Fermi--Dirac lepton-number distributions at $\alpha_{\bar\nu_e}=0.00$, $0.10$, $0.28$, and $0.50$. Left: energy-integrated ELN $G(v)$. A crossing appears for $\alpha_{\bar\nu_e}\gtrsim 0.28$, with $\bar\nu_e$ the more forward species. Right: angle-integrated spectral lepton number $g(\hat{E})$. Different $\beta$ produce an energy-dependent ELN; the angular dial barely changes $g(\hat{E})$.}
    \label{fig:dist_fd}
\end{figure}

The scan is drawn in $(\alpha_{\bar\nu_e},s)$ with isotropic $\nu_e$, chosen so that one angular parameter plays the role that $\eta_F$ played on the ramp. Fixed parameters are $\alpha_{\nu_e}=0$, $\alpha_{\nu_x}=\alpha_{\bar\nu_x}=0.5$, $\beta_{\nu_e}=1.3$, $\beta_{\bar\nu_e}=1.13$, $\beta_x=1$, and number densities locked to $n_{\nu_e}:n_{\bar\nu_e}:n_x=1:0.8:0.4$. The scanned coordinate is $\alpha_{\bar\nu_e}\in[0,\alpha_x]=[0,0.5]$. This is a robustness check of the SFI--FFI transition. Two different $\beta$'s still leave an energy-dependent ELN, unlike the energy-independent occupations of Eqs.~(\ref{eq:nnu})--(\ref{eq:nnubar}). Figure~\ref{fig:dist_fd} shows $G(v)$ and the angle-integrated $g(\hat{E})$ for $\alpha_{\bar\nu_e}=0.00$, $0.10$, $0.28$, and $0.50$. The angular dial crosses $G(v)$ while $g(\hat{E})$ stays nearly the same.

Signed occupancies are the flavor differences of Eq.~(\ref{eq:minerbo}), including the energy-space measure $\hat{E}^2$,
\begin{align}
  g_{\nu}(E,v)
  &\propto \bigl[f_{\nu_e}(E,v)-f_{\nu_x}(E,v)\bigr]\,\hat{E}^2,
  \label{eq:gFD}\\
  g_{\bar\nu}(E,v)
  &\propto -\bigl[f_{\bar\nu_e}(E,v)-f_{\bar\nu_x}(E,v)\bigr]\,\hat{E}^2,
\end{align}
with a single number lock on the three species so that the quoted density ratios are exact. As in Eq.~(\ref{eq:gi}), neutrino bins enter $G_0$ and $G_1$ with a plus sign and antineutrino bins with a minus sign.

\begin{figure}[htbp]
    \centering
    \includegraphics[width=.75\linewidth]{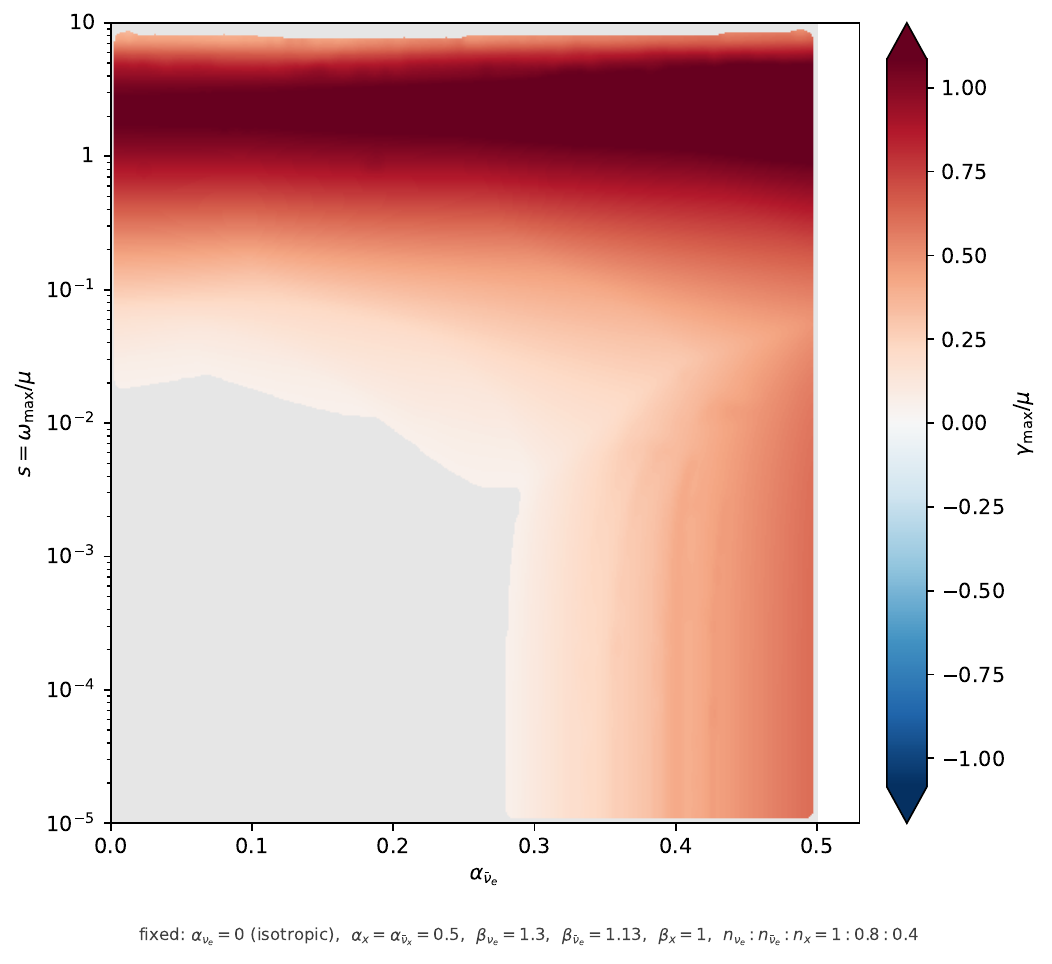}
    \caption{Fermi--Dirac $\gamma_{\max}/\mu$ versus $\alpha_{\bar\nu_e}$ and $s$, $N_v=N_E=256$, from the same linearized problem as Eqs.~(\ref{eq:disp})--(\ref{eq:gmaxkstar}). Fixed: $\alpha_{\nu_e}=0$ (isotropic), $\alpha_x=\alpha_{\bar\nu_x}=0.5$, $\beta_{\nu_e}=1.3$, $\beta_{\bar\nu_e}=1.13$, $\beta_x=1$, $n_{\nu_e}:n_{\bar\nu_e}:n_x=1:0.8:0.4$. Grey: no growing mode. The FFI wall is $\alpha_{\bar\nu_e}\gtrsim 0.28$.}
    \label{fig:gmax_FD}
\end{figure}

Figure~\ref{fig:gmax_FD} is $\gamma_{\max}(\alpha_{\bar\nu_e},s)$ at resolution $N_v=N_E=256$. Grey is a numerical floor $\gamma_{\max}/\mu<0.05$, chosen more conservatively than the ramp floor of $0.02$: no growing root of Eq.~(\ref{eq:disp}) is trusted below it. The FFI wall is $\alpha_{\bar\nu_e}\gtrsim 0.28$: those backgrounds remain unstable down to $s=0$. For smaller $\alpha_{\bar\nu_e}$ the antineutrinos are too isotropic to cross the ELN; the unstable region at large $s$ becomes grey as $s$ drops, which is SFI-only in the same sense as $\eta_F<8/15$ on the ramp.

\begin{figure}[htbp]
    \centering
    \includegraphics[width=.75\linewidth]{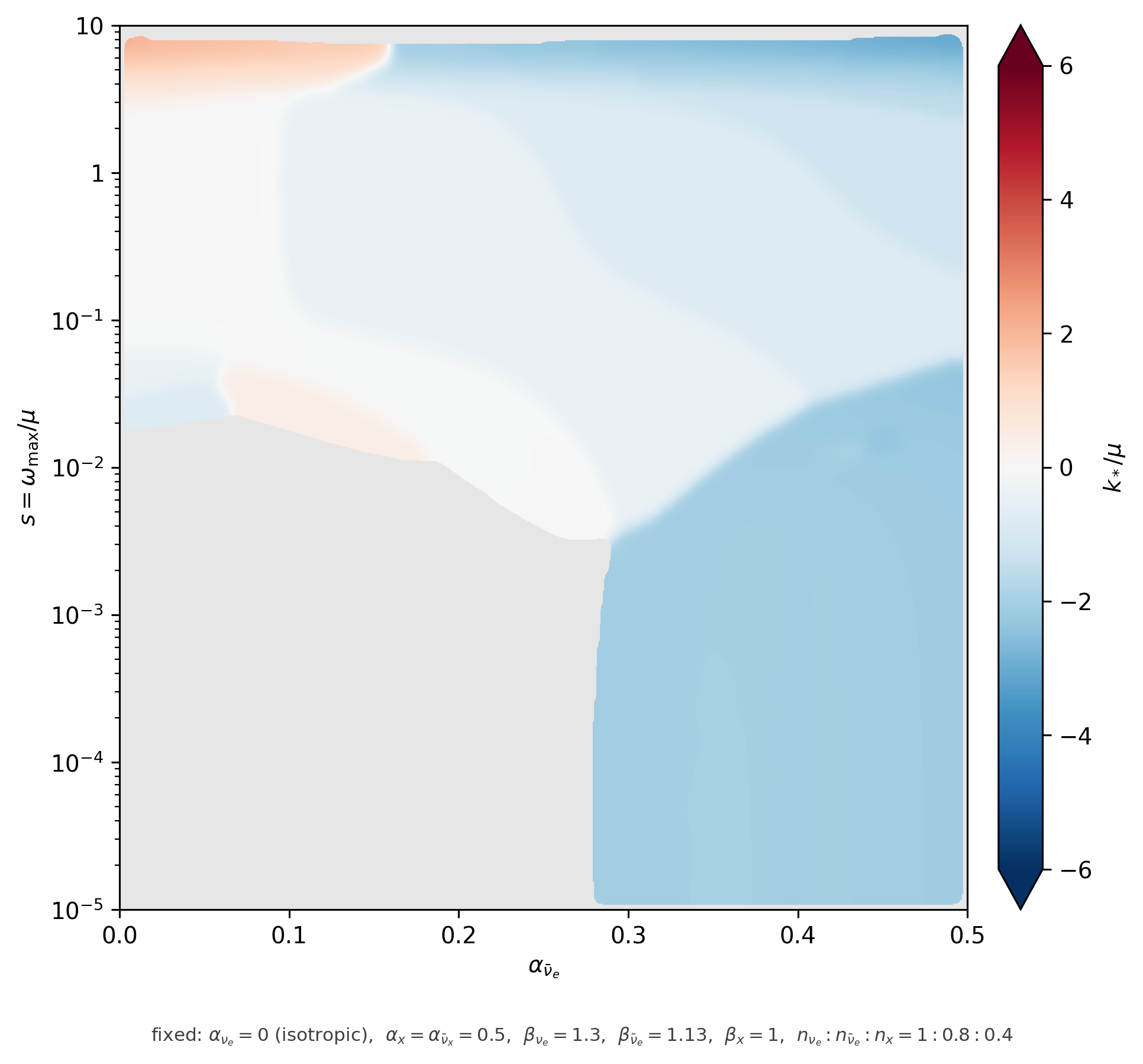}
    \caption{Fermi--Dirac $k_*/\mu$ on the same plane as Fig.~\ref{fig:gmax_FD}. The FFI wall has $k_*<0$ because $\bar\nu_e$ is forward-peaked. Grey: no growing mode.}
    \label{fig:kstar_FD}
\end{figure}

Figure~\ref{fig:kstar_FD} is $k_*$ on the same plane. The FFI wall is blue ($k_*<0$) because $\bar\nu_e$ is the more forward species. That is the same physics as the red FFI wall of Fig.~\ref{fig:kstar_FFISFI} with the opposite beam: the sign of $k_*$ follows which species is more forward-peaked, not the type. At large $s$ the unstable region extends to $\alpha_{\bar\nu_e}=0$ with $k_*$ near zero, consistent with the behavior of SFI in the ramp model.

\begin{figure*}
    \centering
    \includegraphics[width=\linewidth,height=0.42\textheight,keepaspectratio]{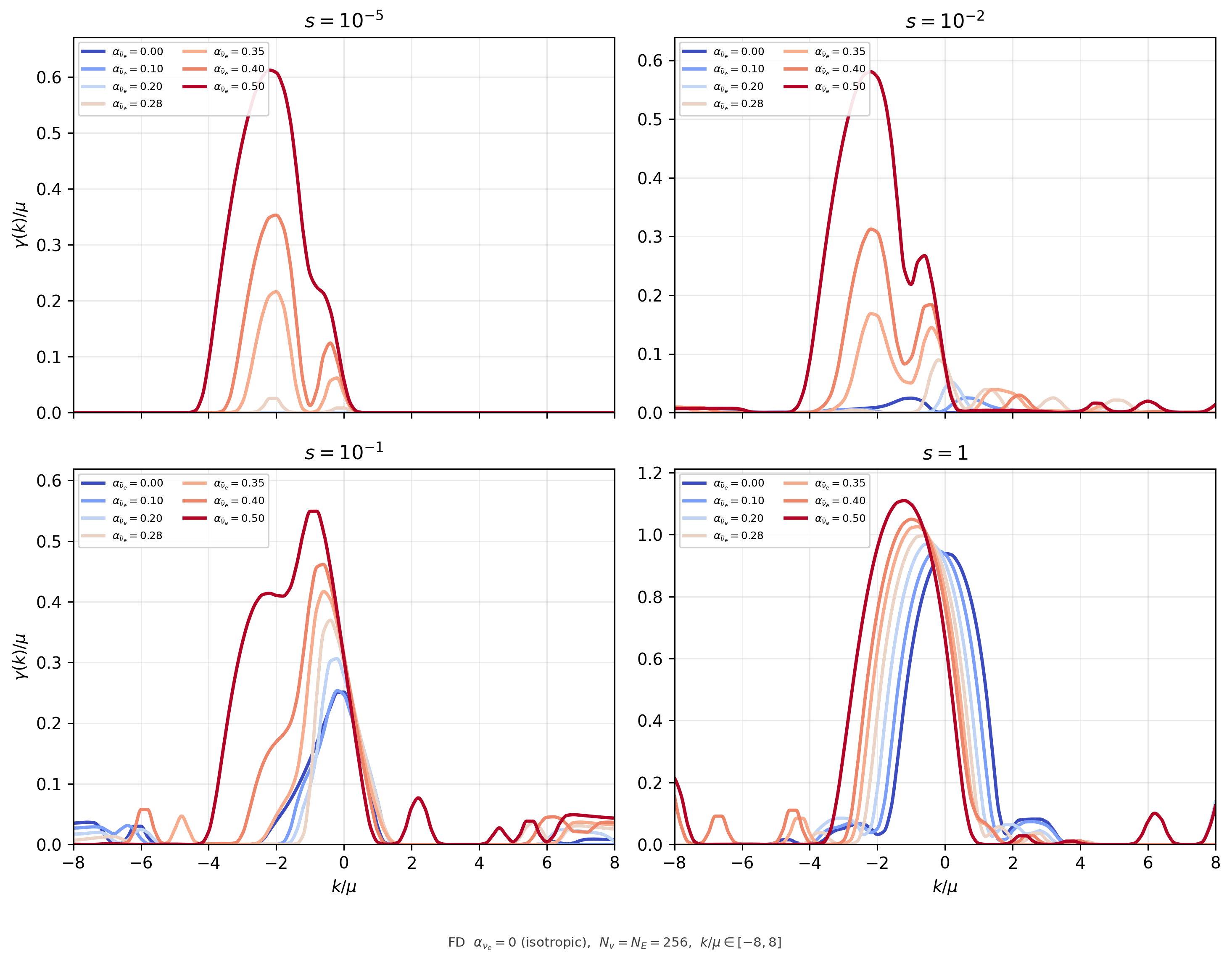}
    \caption{Growth-rate spectra $\gamma_k$ versus $k$ for the Fermi--Dirac family of Figs.~\ref{fig:gmax_FD} and~\ref{fig:kstar_FD}, $N_v=N_E=256$. Each panel is one $s$; curves are $\alpha_{\bar\nu_e}$. FFI (large $\alpha_{\bar\nu_e}$) remains peaked at $k<0$ as $s\to 0$. SFI-only (small $\alpha_{\bar\nu_e}$) is present at $s=1$ and absent at $s=10^{-5}$.}
    \label{fig:gammak_FD}
\end{figure*}

Figure~\ref{fig:gammak_FD} is the Fermi--Dirac analogue of Fig.~\ref{fig:gammak_FFISFI}: the same four values of $s$, curves labeled by $\alpha_{\bar\nu_e}$. At $s=10^{-5}$ only $\alpha_{\bar\nu_e}\gtrsim 0.28$ grows, peaked at $k<0$, with $\gamma_0$ small: an FFI flavor wave. At $s=10^{-2}$ the wall remains and small $\alpha_{\bar\nu_e}$ still has no growing mode. At $s=10^{-1}$ and $s=1$ every $\alpha_{\bar\nu_e}$ in the scan is unstable. The left-hand curves that vanish as $s$ is lowered are SFI-only in the same sense as $\eta_F=0.20$ on the ramp.

\section{Nonlinear evolution of the wall}
\label{sec:nl}

Linear analysis of Eq.~(\ref{eq:qke}) answers a question about a tiny seed. A Fourier mode of flavor coherence grows as $e^{\lambda t+ikz}$ with $\lambda=\gamma+i\Omega$ from Eqs.~(\ref{eq:ansatz})--(\ref{eq:lambda}). The spectrum $\gamma_k$ of Eq.~(\ref{eq:gammak}) is the exponential growth rate of each wavenumber; $k_*$ of Eq.~(\ref{eq:gmaxkstar}) is the wavenumber of fastest growth. On the SFI-only side of the wall, $\gamma_k$ is unimodal and peaked near $k=0$: the seed is a spatially nearly uniform mode that needs vacuum frequency. On the FFI side, $\gamma_k$ is peaked at finite $k$ with $\gamma_0\simeq 0$: a band of flavor waves grows, not a single mode, and they do not need vacuum frequency. That is all the linear problem can say. Once the coherence is no longer small, $\gamma_k$ is no longer the evolution. The same collisionless QKE must be integrated, and the question is whether the gas keeps the spatial structure of the linear peak.

We evolve four backgrounds, the two sides of the wall on each family, in a one-dimensional periodic box of length $L=40\pi$ with $N_x=512$ spatial cells, the same geometry used in collisionless FFI studies~\cite{wu2021collective,richers2022code}. The two ramp points are $(\eta_F,s)=(0.20,1)$ and $(0.70,10^{-3})$ on the map of Fig.~\ref{fig:gmax_FFISFI}. The two Fermi--Dirac points are $\alpha_{\bar\nu_e}=0.10$ at $s=1$ and $\alpha_{\bar\nu_e}=0.50$ at $s=10^{-3}$ on Fig.~\ref{fig:gmax_FD}. The angular/energy grids match the linear maps: $N_v=N_\omega=256$ on the ramp (occupancy rescaled by $8/N_\omega$ as in Sec.~\ref{sec:methods}) and $N_v=N_E=256$ on the Fermi--Dirac family. The stepper is a collisionless fourth-order Runge--Kutta (RK4) integration of Eq.~(\ref{eq:qke}) with an off-diagonal seed of amplitude $10^{-8}$. Each run covers $t\mu=800$.

The diagnostic is the same Fourier content that $\gamma_k$ diagnosed while the seed was small. The flavor coherence of mode $i$ is the off-diagonal element $S_i(x,t)\equiv\rho_{i,ex}(x,t)$ of Eq.~(\ref{eq:qke}). Its discrete Fourier amplitude on the periodic domain is
\begin{equation}
  \tilde S_i(k,t)
  =\frac{1}{N_x}\sum_x e^{-ikx}\,S_i(x,t).
  \label{eq:Stilde}
\end{equation}
Summing in quadrature over all momentum bins $i$ (angle, energy, and species) gives the amplitude of flavor waves at wavenumber $k$,
\begin{equation}
  A(k,t)
  =\Biggl(\sum_i\bigl|\tilde S_i(k,t)\bigr|^2\Biggr)^{1/2},
  \label{eq:Ak}
\end{equation}
and the power spectrum
\begin{equation}
  P(k,t)=A(k,t)^2.
  \label{eq:Pkt}
\end{equation}
$P(k,t)$ is the nonlinear counterpart of $\gamma_k$: it is the actual power in flavor coherence at each $k$ after the amplitude has left the linear regime~\cite{richers2022code,richers2021neutrino}. A peak of $P$ around $k=0$ is a nearly homogeneous rearrangement of flavor~\cite{hannestad2006self,padillagay2025flavor}. A peak of $P$ at finite $k$ is a flavor wave of wavelength $2\pi/k$~\cite{yi2019dispersion, duan2021flavor, fiorillo2025collective, fiorillo2026flavomon, fiorillo2026quasi, johns2025local, kost2026neutrino, liu2026dynamical}. If the linear classification survives, that peak sits at the linear $k_*$.

\begin{figure*}
    \centering
    \includegraphics[width=\linewidth]{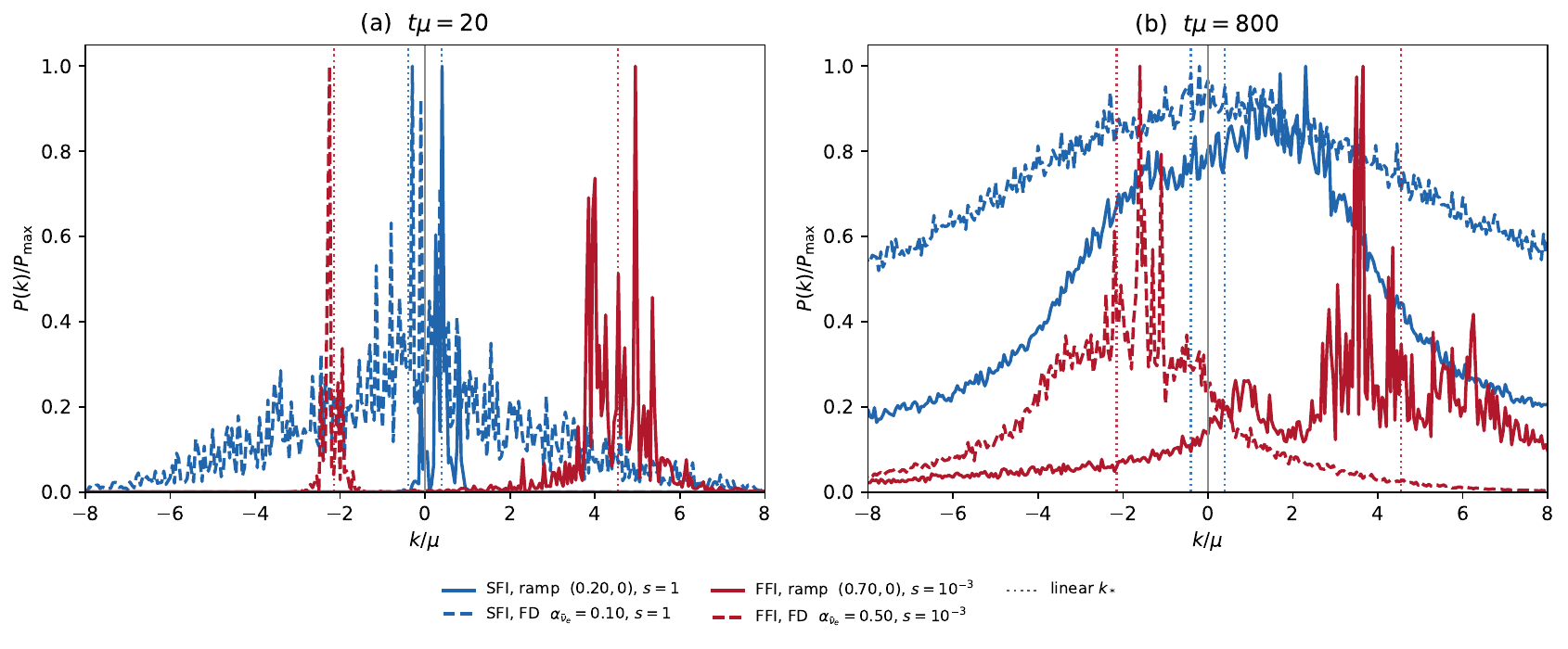}
    \caption{Flavor-wave spectrum $P(k)$ of Eq.~(\ref{eq:Pkt}), normalized to its maximum, on both sides of the wall and on both families. $N_x=512$. Solid: ramp; dashed: Fermi--Dirac. Blue: large $s$, uncrossed, peaked near $k=0$. Red: small $s$, crossed, peaked at finite $k$. Dotted: linear $k_*$. (a)~$t\mu=20$, still close to the linear peak. (b)~$t\mu=800$, after saturation.}
    \label{fig:nl_P}
\end{figure*}

Figure~\ref{fig:nl_P} shows $P(k)$ for all four models: ramp solid, Fermi--Dirac dashed, large-$s$ uncrossed side blue, small-$s$ crossed side red; dotted lines are the linear $k_*$. At $t\mu=20$, while the amplitude is still close to the linear problem, power on the uncrossed side is peaked near $k=0$ and power on the crossed side is peaked at finite $k$, near the linear $k_*$ of each model. At $t\mu=800$ the spectra are broader, but that split remains: the first pair stays a band around $k=0$, the second a band at finite $k$. Solid and dashed curves of the same side overlie, so the occupancy formula is not what sets the late-time $k$ content. Opposite signs of $k_*$ on the crossed side follow which species is more forward-peaked.

\begin{figure}[htbp]
    \centering
    \includegraphics[width=.95\linewidth]{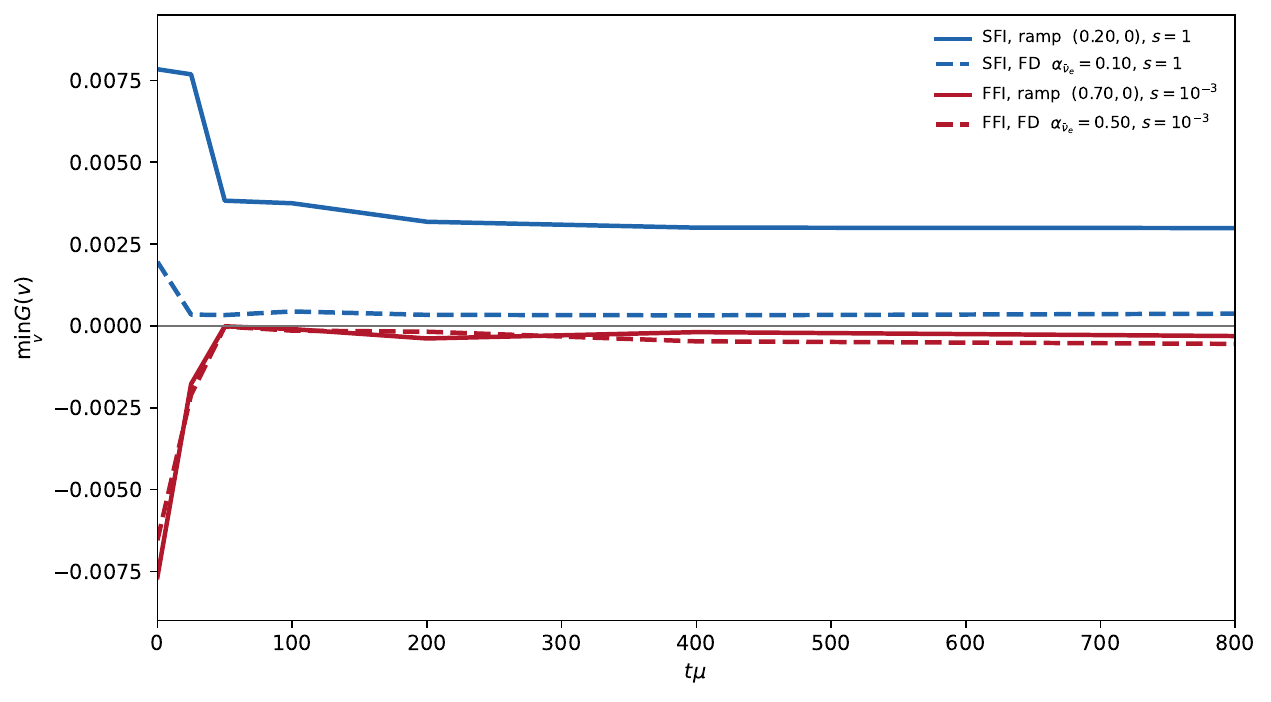}
    \caption{Same four models as Fig.~\ref{fig:nl_P}. $N_x=512$. Solid: ramp; dashed: Fermi--Dirac. The $y$-axis is $\min_v G(v)$ of the energy-integrated ELN, Eq.~(\ref{eq:Gv}): an angular crossing exists iff this number is negative. Blue: large $s$, uncrossed, stays positive. Red: small $s$, crossed, starts negative and is washed to $\simeq 0$.}
    \label{fig:nl_eln}
\end{figure}

Figure~\ref{fig:nl_eln} is the complementary half of the same transition. Here $G(v)$ is the energy-integrated ELN of Eq.~(\ref{eq:Gv}): an angular crossing exists if and only if $\min_v G(v)<0$. That is the $y$-axis. Both SFI curves stay positive: no crossing is created. Both FFI curves start negative and are washed to $\min G\simeq 0$ by $t\mu\sim 50$, the collisionless FFI flattening that removes the crossing which caused the instability~\cite{bhattacharyya2020late,zaizen2023simple,richers2024asymptotic,liu2024quasi}. Ramp and Fermi--Dirac of the same side overlie. Figure~\ref{fig:nl_P} says where the flavor power sits; Fig.~\ref{fig:nl_eln} says whether a crossing exists. That pair is the nonlinear content of the transition.

These four models represent the two sides of the ELN-crossing divide ($\eta_F=8/15$ on the ramp), not a scan through the divide itself. Linearly, $\gamma_k$ selects a wavenumber. Nonlinearly, $P(k)$ says where the flavor power sits and $\min_v G$ says whether an ELN crossing exists. SFI: power around $k=0$, no crossing. FFI: flavor wave that has flattened its own crossing. Those are different inputs to neutrino transport.

\section{Discussion}
\label{sec:discussion}

The object of this paper is mixed-type instabilities away from the idealized slow and fast limits. An instability that still grows as $\omega\to 0$ is fast in the sense of Sawyer~\cite{sawyer2005speed,sawyer2016neutrino}. Slow instabilities can be seen as modes that disappear at $\omega=0$, or as modes that persist when the angular distributions are restored to isotropy. Away from those two limits, SFI and FFI are blended; they can be projected onto either limit by setting $\omega=0$, by isotropizing, or both.

On the ramp plane $(\eta_F,s)$, Figs.~\ref{fig:gmax_FFISFI} and~\ref{fig:kstar_FFISFI}, growth at large $s$ on the uncrossed side of $G(v)$ is gone as $s\to 0$, while for $\eta_F\gtrsim 8/15$ it continues to $s=0$. The sign of $k_*$ follows which species is more forward-peaked, and $\gamma_0/\gamma_{\max}$ says whether the fastest mode is spatially uniform or a flavor wave; neither is an SFI/FFI classifier. Figure~\ref{fig:gammak_FFISFI} shows a smooth crossover of $k_*$ while $\gamma_{\max}$ stays smooth: the location of the maximum of $\gamma_k$ moves from near $k=0$ to finite $k$. Fermi--Dirac occupations, Figs.~\ref{fig:gmax_FD}--\ref{fig:gammak_FD}, show the same structure with the opposite $k_*$ sign because antineutrinos are the more forward species.

That is not the same as the existence of a mixed energy--angle crossing. Recent analyses~\cite{dasgupta2025sufficient,fiorillo2025dispersion} ask whether unstable modes exist when both energy and angle enter the lepton-number distribution. A unifying dispersion relation can house slow, fast, and collisional roots in one linearized problem~\cite{fiorillo2025dispersion}. Neither construction maps mixed-type instabilities away from those idealized limits. In the reduced parameter spaces of Figs.~\ref{fig:gmax_FFISFI} and~\ref{fig:gmax_FD} a weaker growing branch can be present alongside the fastest mode.

Nonlinear evolution of both sides of the ELN-crossing divide, in a collisionless one-dimensional periodic box with $N_x=512$, is Figs.~\ref{fig:nl_P} and~\ref{fig:nl_eln}~\cite{johns2020fast,wu2021collective,richers2022code}. On the uncrossed, large-$s$ side, $P(k)$ stays a band around $k=0$ and $\min_v G$ stays positive. On the crossed, small-$s$ side, $P(k)$ stays a band at finite $k$ and the energy-integrated ELN crossing is washed, consistent with collisionless FFI flattening and with the asymptotic-state picture in which the instability removes its own ELN crossing~\cite{bhattacharyya2020late,zaizen2023simple,richers2024asymptotic,liu2024quasi}. Ramp and Fermi--Dirac of the same side overlie.

Those two late-time states are different inputs to neutrino transport. Subgrid closures built around a local instability type~\cite{johns2024subgrid,xiong2024robust,ehring2023fast} do not treat a nearly homogeneous rearrangement and a finite-$k$ wave that has flattened $G(v)$ as the same object. A diagnosis that uses only the FFI limit---an ELN crossing at $\omega=0$~\cite{wu2017fast,li2021neutrino,george2020fast,grohs2023neutrino}---is incomplete once vacuum frequencies are not negligible.

The remaining pure type is collisional. The SFI--CFI and FFI--CFI parameter spaces are a natural next step toward flavor instabilities as they occur in compact-object environments.

\begin{acknowledgments}
NV acknowledges support from the Millennium
Institute of Subatomic Physics at High Energy Frontier
ICN2019 044. LJ is supported by a Feynman Fellowship through the Laboratory Directed Research and Development (LDRD) program at Los Alamos National Laboratory, project number 20230788PRD1.
\end{acknowledgments}

\bibliography{refs_v3}

\end{document}